\documentclass[oneside]{article}
\usepackage{lmodern}
\usepackage[T1]{fontenc}
\usepackage[latin9]{inputenc}
\usepackage{textcomp}
\usepackage{mathtools}
\usepackage{amsmath}
\usepackage{graphicx}
\usepackage[english]{babel}
\usepackage{authblk}
\usepackage{float}
\usepackage{xfrac}
\usepackage{esint}
\usepackage[unicode=true,
 bookmarks=false,
breaklinks=false,pdfborder={0 0 1},backref=false,colorlinks=false]
 {hyperref}
\usepackage[a4paper, total={210mm,297mm},margin=2.0cm]{geometry}

\usepackage{datetime}
\newdate{date}{29}{2}{2016}

\providecommand{\tabularnewline}{\\}

\title{Three-Dimensional Model for Electrospinning Processes in Controlled Gas Counterflow}

\author[1]{Marco Lauricella}
\author[2,3]{Dario Pisignano}
\author[1,4]{Sauro Succi \thanks{Electronic address: \texttt{s.succi@iac.cnr.it}; Corresponding author}}

\affil[1]{Istituto per le Applicazioni del Calcolo CNR, Via dei Taurini 19, 00185 Rome, Italy}
\affil[2]{Dipartimento di Matematica e Fisica "Ennio De Giorgi", University of Salento, via Arnesano, 73100 Lecce, Italy}
\affil[3]{Istituto Nanoscienze-CNR, Euromediterranean Center for Nanomaterial Modelling and Technology (ECMT), via Arnesano, 73100 Lecce, Italy}
\affil[4]{Harvard Institute for Applied Computational Science, Cambridge, Massachusetts, USA}

\date{\displaydate{date}}

\begin{document}

\maketitle

\textbf{This is an unofficial adaptation of an article that appeared in an ACS publication. ACS has not endorsed the content of this adaptation or the context of its use.}

\begin{abstract}
We study the effects of a controlled gas flow on the dynamics of electrified
jets in the electrospinning process. The main idea is to model the
air drag effects of the gas flow by using a non-linear Langevin-like
approach. The model is employed to investigate the dynamics of electrified
polymer jets at different conditions of air drag force, showing that
a controlled gas counterflow can lead to a decrease of the average
diameter of electrospun fibers, and potentially to an improvement
of the quality of electrospun products. We probe the influence of
air drag effects on the bending instabilities of the jet and on its
angular fluctuations during the process. The insights provided by this study might
prove useful for the design of future electrospinning experiments and
polymer nanofiber materials. 
\end{abstract}

\section{Introduction}

The production of nano- and microfibers has gained increasing interest
due to the large number of promising applications, including filtration,
textiles, medical, protective, structural, electrical, and optical
materials and coatings. In particular, an intriguing feature of
electrospun fibers is the high surface-area, which is due to the combination of small radius
and extreme length of the fiber (in principle
up to km when polymer solutions with high degree of molecular entanglement
are used to achieve stable electrified jets). This offers intriguing perspectives for practical applications.
As a consequence, several studies have been focused on the production
and characterization of such structures. \cite{reneker1996nanometre,li2004electrospinning2,ramakrishna2005introduction,luo2012electrospinning,wendorff2012electrospinning,pisignanoelectrospinning,persano2015active} 

Following the pioneering works of Rayleigh \cite{rayleigh1882equilibrium}
and, later, Zeleny,\cite{zeleny1917instability} the electrospinning
process relies on a strong electric field (typically $10^{5}-10^{6}\text{ V}\cdot\text{m}^{-1}$)
in order to elongate and accelerate a polymeric fluid body from a
nozzle towards a conductive collector. During the development of the jet path, the stream
cross-section decreases by orders of magnitude, providing a jet, and
consequently solid fibers, with transversal size potentially well
below the micrometer-scale. The dynamic evolution of the 
polymer nanojet involves two different stages: in the first, the pendent polymeric droplet is
stretched by the intense external electric field, providing a straight path. 
In the second, small perturbations induce bending instabilities, and a complex jet path is
consequently observed. In a typical electrospinning
experiment, hydrodynamic perturbations, as well as mechanical vibrations
nearby the nozzle, might misalign the jet axis. According to the Earnshaw's
theorem,\cite{jeans1908mathematical} an off-axis misalignment triggers
an electrostatic-driven bending instability, leading the fluid into
a region of spiral coils. As a consequence, the jet travels a larger
distance between the nozzle and the collector, and the fiber diameter
undergoes a further decrease along the way, leading to a reduction
of the fiber diameter.

Several studies were focused on experimental parameters, such as applied
electric voltage, liquid viscosity, etc..\cite{fong1999beaded,theron2004experimental,carroll2006electrospinning,montinaro2015dynamics}
Similarly, the use of complementary external forces was also investigated.
For instance, a gas stream provided by suitable distributers and surrounding
the electrospinning nozzle can be used as additional stretching force,
providing fibers with small diameter. \cite{wang2005formation,yao2006polysulfone,kim2008direct,lin2008preparation,zhmayev2010nanofibers,hsiao2012effect}
This process is generally called gas-assisted electrospinning (sometimes
electroblowing). Nonetheless, many of the effects of gas flows on
electrospinning still need to be investigated in a systematic way,
particularly with regard to the relationship between gas flow speed
and bending instabilities. Indeed, given the ubiquitous nature of
intentional or stochastic gas flows in the process atmosphere, understanding
in depth such points is very important for a correct design of electrospinning
experiments, when fibers with very small diameters are to be produced
with a given polymer solution.

In this framework, simulation models can be useful for understanding
the key processing parameters and ultimately exerting a better
control on the resulting fiber morphologies, better elucidating the
phenomenology of electrified jets and providing valuable information
for the development of new spinning experiments. For
these reasons, various models have been proposed for electrospinning
in the recent years,\cite{reneker2000bending,yarin2001taylor,hohman2001electrospinning,fridrikh2003controlling,theron2004experimental,carroll2006electrospinning}
which can be categorized on based on the approach used for representing
the jet. In the first class of models, the filament is treated as
obeying the equations of continuum mechanics,\cite{spivak2000model,feng2002stretching,feng2003stretching,hohman2001electrospinning,hohman2001applications}
whereas in the second the jet is described as a series of discrete
elements obeying the equations of Newtonian mechanics,\cite{reneker2000bending,yarin2001taylor}
as it is the case of the present work.

Recently, Lauricella et al. \cite{lauricella2015nonlinear} developed
a one-dimensional model for studying the air drag effects on the early
stage of electrospinning process. In this approach, the liquid jet
was represented as a series of charged beads, connected by viscoelastic
springs according to the original picture proposed by Reneker and
Yarin.\cite{reneker2000bending,yarin2001taylor} The jet dynamics
was the result of the combined action of viscoelastic Coulombic, external
electrical forces, and a dissipative term which models the air drag
effect. Based on experimental observations,\cite{spinning1991science}
the dissipative air drag term was taken as non-lineary dependent on
the jet geometry. As consequence, the model included a non-linear
Langevin-like stochastic differential equation describing the fluid
motion. However, an investigation of the air drag effects on three-dimensional
(3D) bending instabilities was still missing.

Here, we provide a 3D description of electrified jets which includes
air drag, and study its effects in the dynamics of the bending instabilities.
In particular, our aim is to investigate the relation between the dissipative-perturbing
forces and the resulting deposition of electrospun fibers. Furthermore,
the extended model is used to set up an ideal experiment of gas-assisted
electrospinning, which involves a gas-injecting system located at
the collector and oriented towards the spinneret. In this context,
we probe the effects of a controlled gas counterflow on the fiber
diameter, which could be useful for designing new electrospinning
experiments.

The article is organized as follows. In Sec. 2, we present the 3D
model for electrospinning, with the set of stochastic differential
equations of motion (EOM) which govern the dynamics of system. Results
are reported and discussed in Sec. 3. Finally, conclusions are outlined
in Sec. 4.

\section{Model of electrospinning in a gas flow}

\begin{figure}[H]
\begin{centering}
\includegraphics[scale=0.5]{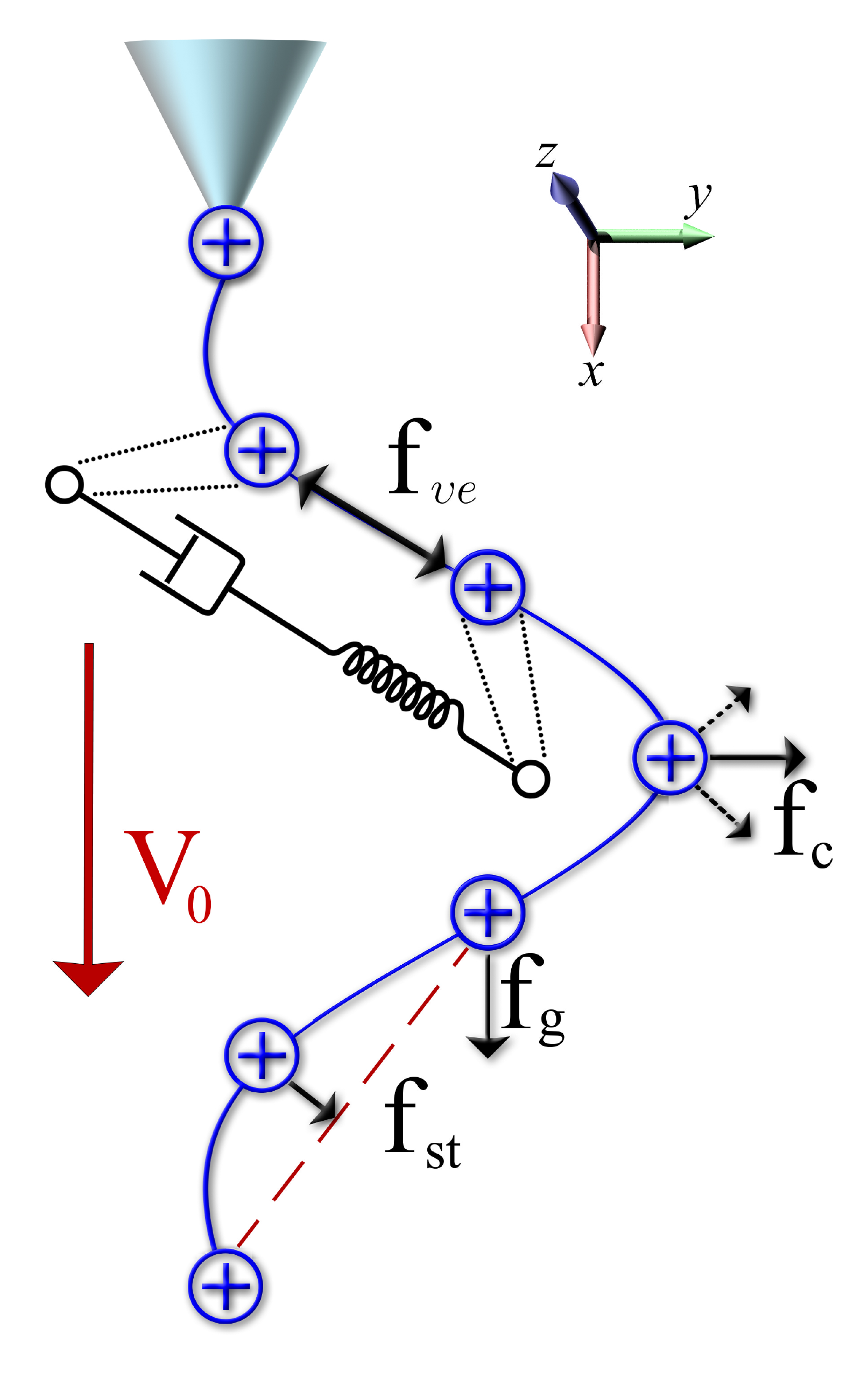} 
\par\end{centering}

\protect\caption{Diagram of the electrospinning model as implemented 
in the "vanilla" version of JETSPIN without air drag 
and lift force terms (which are sketched in Fig \ref{Fig:lift-drag-forces}). 
Each discrete element representing a jet segment is drawn
by a blue circle with a plus sign denoting the positive charge of
segment. We represent the Maxwell viscoelastic force, $\textbf{f}_{\upsilon e}$, 
the gravitational force $\vec{\textbf{f}}_{g}$,
the surface tension force, $\textbf{f}_{st}$, pointing the center
of curvature to restore the rectilinear shape, and the Coulomb repulsive
term, $\textbf{f}_{c}$, which is the sum over all the repulsive interactions
between the beads. The external electric potential, $V_{0}$, is indicated
by the red arrow in figure, while the upper cyan cone represents the
nozzle. The dashed red line represents the ideal straight line to which
the filament tends under the surface tension force.}

\label{Fig:model} 
\end{figure}

In this paper, we modify the 3D model of electrospinning previously implemented
in the software package JETSPIN, a specifically-developed, open-source
and freely-available code.\cite{lauricella2015jetspin,jetspin} 
We use a Lagrangian discrete model which represents the
polymer solution filament as a series of $n$ beads (jet beads) at
mutual distance $l$, each pair of beads in the row being connected
by viscoelastic elements, as proposed in Ref. \cite{reneker2000bending}
(Fig. \ref{Fig:model}). The length $l$ is taken to be larger than
the radius of the filament.
Each $i-th$ bead has mass $m_{i}$ and charge $q_{i}$, assumed equal for all the beads for simplicity. 
The spinneret is represented
by a single mass-less point of charge $q_{0}$ fixed at $x=0$, which
we call nozzle bead. A typical simulation is started with a single
jet bead inserted at the nozzle, and placed at distance $l_{step}$
from the nozzle along the $x$ axis. The onset of the jet takes place
with a cross-sectional radius $a_{0}$, defined as the radius of the
polymer solution filament at the nozzle, before the stretching process
occurs, leading to the elongation and cross-section reduction in the
fluid body. Furthermore, the starting jet bead has an initial velocity
$\upsilon_{s}$ along the $x$ axis equal to the bulk fluid velocity
in the needle of the extrusion syringe or reservoir. Once this traveling
bead reaches a distance $2\cdot l_{step}$ away from the nozzle, a new
particle (third body) is placed at distance $l_{step}$ from the nozzle
along the straight line joining the two previous bodies (nozzle and
previous jet bead).
Note that $l_{step}$ defines the length step used to discretize 
the liquid jet at the nozzle before the stretching process starts taking place.
The procedure is then repeated, leading to
a series of $n$ beads representing the jet. 
It is worth stressing
that hereafter we indicate by $i=1$ the particle which
is the closest to the collector. 

The jet is therefore modelled as a body constituted by a viscoelastic
Maxwell fluid, and the stress $\sigma_{i}$ on the $i-th$ dumbbell
which connects the bead $i$ with the bead $i+1$ is given by the
equation:

\begin{equation}
\frac{d\sigma_{i}}{dt}=\frac{G}{l_{i}}\frac{dl_{i}}{dt}-\frac{G}{\mu}\sigma_{i},\label{eq:stress-ode}
\end{equation}
where $l_{i}$ is the length of the element, $G$ is the elastic modulus,
$\mu$ the viscosity of the fluid jet, and $t$ the time (see Fig.
\ref{Fig:model}). The length $l_i$ is computed as 
the mutual distance between the $i-th$ bead and its previous bead.
Being $a_{i}$ the fiber radius 
at the bead $i$, the viscoelastic force, $\vec{\textbf{f}}_{\upsilon e}$,
pulling the bead $i$ back to $i-1$ and towards $i+1$, reads as
follows:

\begin{equation}
\vec{\textbf{f}}_{\upsilon e,i}=-\pi a_{i}^{2}\sigma_{i}\cdot\vec{\textbf{t}}_{i}+\pi a_{i+1}^{2}\sigma_{i+1}\cdot\vec{\textbf{t}}_{i+1}\text{,}
\end{equation}
where $\vec{\textbf{t}}_{i}$ is the unit vector pointing from bead
$i-1$ to bead $i$. The force $\vec{\textbf{f}}_{st}$ due to the surface tension
for the $i-th$ bead is given by:

\begin{equation}
\vec{\textbf{f}}_{st,i}=\alpha\,k_{i}\cdot\pi\left(\frac{a_{i}+a_{i-1}}{2}\right)^{2}\cdot\vec{\textbf{c}}_{i}\text{,}
\end{equation}
where $\alpha$ is the surface tension coefficient, $k_{i}$ is the
local curvature, and $\vec{\textbf{c}}_{i}$ is the unit vector pointing
the center of the local curvature from bead $i$ (see Fig. \ref{Fig:model}).
The force $\vec{\textbf{f}}_{st}$ tends to restore the rectilinear
shape acting on the bent part of the jet.

In electrospinning processes, the jet stretch is mainly due to an external electric potential
$V_{0}$ which is applied between the spinneret and the conducting collector.
Denoted by $h$ the distance of the collector from the injection point,
each $i-th$ bead undergoes the electric force:

\begin{equation}
\vec{\textbf{f}}_{el,i}=e_{i}\frac{V_{0}}{h}\cdot\vec{\textbf{x}}\text{,}
\end{equation}
where $\vec{\textbf{x}}$ is the unit vector pointing the collector
from the spinneret (see Fig. \ref{Fig:model}).
Note that whenever a jet bead touches the collector, 
its position is frozen and its charge is set to zero.

The net Coulomb force $\vec{\textbf{f}}_{c}$ on the $i-th$ bead
from all the other beads is given by:

\begin{equation}
\vec{\textbf{f}}_{c,i}=\sum_{\substack{j=1\\
j\neq i
}
}^{n}\vec{f}_{c,i,j}=\sum_{\substack{j=1\\
j\neq i
}
}^{n}\frac{q_{i}q_{j}}{R_{ij}^{2}}\cdot\vec{\textbf{u}}_{ij}\text{,}
\end{equation}
where $R_{ij}^{2}=\left(x_{i}-x_{j}\right)^{2}+\left(y_{i}-y_{j}\right)^{2}+\left(z_{i}-z_{j}\right)^{2}$,
and $\vec{\textbf{u}}_{ij}$ is the unit vector pointing the $i-th$
bead from the $j-th$ bead.

The force due to the gravity is also considered in the model, and it is computed by
the usual expression

\begin{equation}
\vec{\textbf{f}}_{g,i}=m_{i}g\cdot\vec{\textbf{x}},
\end{equation}
where $g$ is the gravitational acceleration.

These features are implemented in the JETSPIN software package.\cite{lauricella2015jetspin}
Next, we extend the 3D framework in order to include the air drag
terms, and reproduce aerodynamic effects. Consequently, code modifications
have been implemented in JETSPIN. In particular, we model the air
drag by adding a random term and a dissipative term to the forces involved
in the process. The dissipative air drag term is usually dependent
on the geometry of the jet, which changes in time, and it combines
longitudinal and lateral components. Based on experimental findings,\cite{spinning1991science,sinha2010meltblowing,yarin2014fundamentals}
the longitudinal component of the air drag dissipative force term
acting on a jet segment of length $l$ is given by the empirical formula:

\begin{equation}
\vec{\textbf{f}}_{air}=l\cdot0.65\pi a\rho_{a}\upsilon^{2}_{t}\left(\frac{2\upsilon_{t} a}{\nu_{a}}\right)^{-0.81} 
\cdot\vec{\textbf{t}} ,
\label{eq:fric-emp}
\end{equation}
where $\rho_{a}$ denotes the air density, $\nu_{a}$ the kinematic
viscosity, $\vec{\textbf{t}}$ the tangent unit vector, and $\upsilon_{t}=\left(\vec{\boldsymbol{\upsilon}} -\vec{\boldsymbol{\upsilon}}_{flow}\right)\cdot\vec{\textbf{t}}$ 
represents the tangent component of the total velocity with respect to the air flow given as the difference between  jet velocity $\upsilon$ and air flow velocity $\upsilon_{flow}$.
The gas flow is assumed to be oriented along the $x$-axis with opposite direction, but the choice is not mandatory.
Following the approach introduced by Lauricella et al. \cite{lauricella2015nonlinear}, we rearrange the last Eq. as

\begin{equation}
\vec{\textbf{f}}_{air}=l\cdot0.65\pi\rho_{a}\left(\frac{2}{\nu_{a}}\right)^{-0.81}a^{0.19}\upsilon^{1.19}_{t} \cdot\vec{\textbf{t}}.
\label{eq:fric-rewritten}
\end{equation}
Rewriting Eq. \ref{eq:fric-rewritten} for the $i-th$ bead 
representing a jet segment, and assuming a constant volume 
of the jet $\pi a^{2}_i l_i=\pi a_{0}^{2}l_{step}$  , so that
\begin{equation}
 a_i=a_{0}\sqrt{l_{step}/l_i},  
\label{eq:vol-cons}
\end{equation}
with $l_{step}$ and $a_{0}$ respectively the length and 
the radius of the jet segment between
at the nozzle before the stretching, we obtain

\begin{equation}
\vec{\textbf{f}}_{air,i}=-m_{i}\gamma_{i}\,l_{i}^{0.905}\upsilon_{t,i}^{1+0.19}\cdot\vec{\textbf{t}}_{i-1}
\end{equation}
where we have collected several terms of the empirical relationship in $\gamma_{i}$ which is equal to
\begin{equation}
\gamma_{i}=0.65\pi\frac{\rho_{a}}{m_{i}}\left(\frac{2}{\nu_{a}}\right)^{-0.81}l_{step}^{0.095}a_{0}^{0.19},
\label{eq:fric-coef}
\end{equation}
in order to obtain the dissipation term of a non linear Langevin-like equation (for further details see Lauricella et al. \cite{lauricella2015nonlinear}). 
It is worth stressing that $\gamma_{i}$ is derived by the 
empirical relationship of Eq. \ref{eq:fric-emp}, so that 
also Eq. \ref{eq:fric-coef}  is a nondimensional combination 
of physical parameters.

\begin{figure}[H]
\begin{centering}
\includegraphics[scale=0.5]{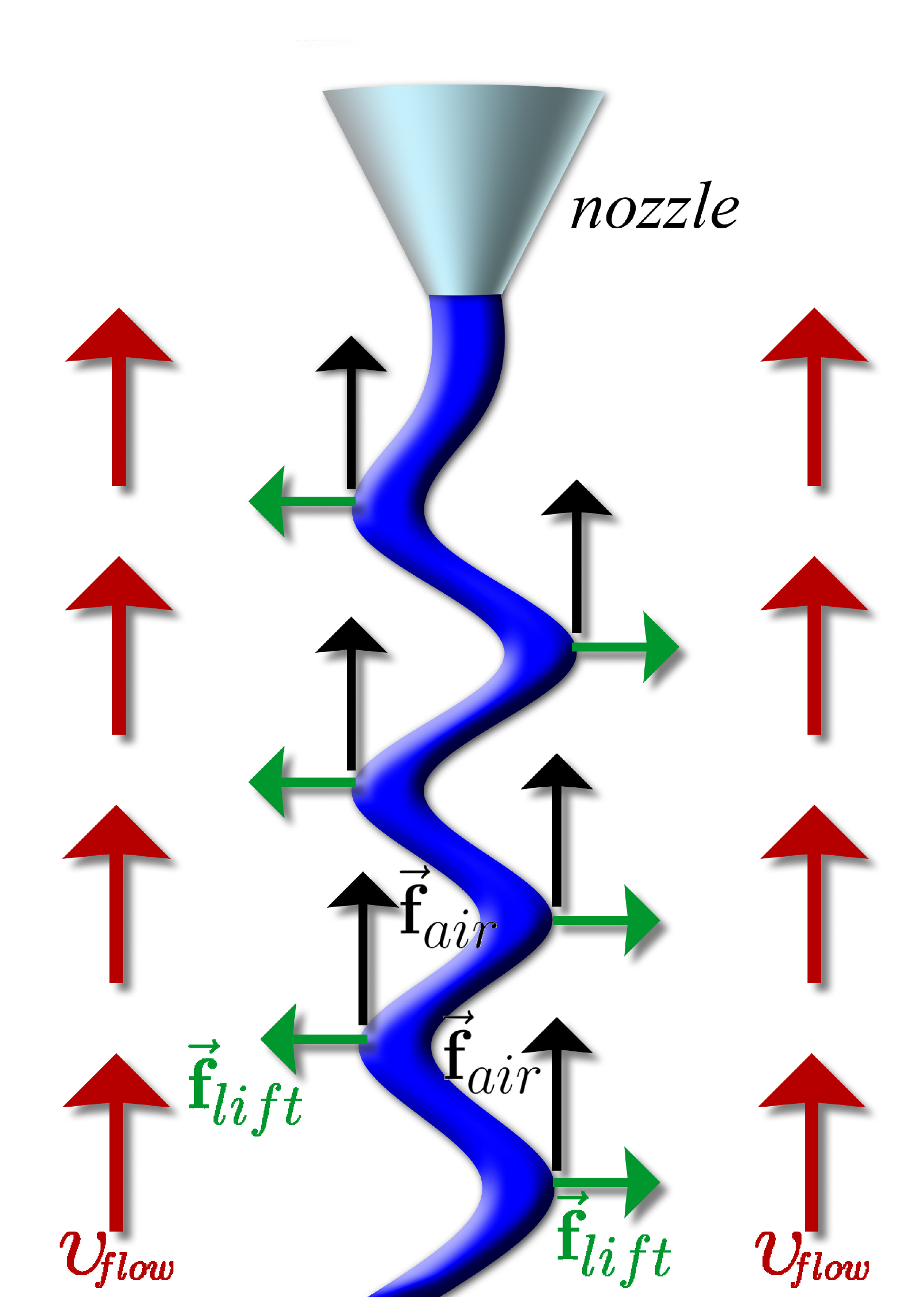} 
\par\end{centering}

\protect\caption{Diagram of the electrospinning model showing the dissipative force,
which is the sum of air drag force, $\vec{\textbf{f}}_{air}$, (black
arrows) and lift force, $\vec{\textbf{f}}_{lift}$, (green arrows),
when a gas flow of speed $\upsilon_{flow}$ is present (red arrows).}

\label{Fig:lift-drag-forces} 
\end{figure}

In a 3D framework a lateral lift force should also be considered.
Following the expression introduced by Yarin,\cite{yarin1993free,yarin2014fundamentals}
under a high-speed air drag the lateral component $\vec{\textbf{f}}_{lift,i}$
of the aerodynamic dissipative force related to the flow speed is
given in the linear approximation (for small bending perturbations)
by: 
\begin{equation}
\vec{\textbf{f}}_{lift,i}=-l_{i}\cdot k_{i} \rho_{a}\upsilon_{t,i}^{2}\pi\left(\frac{a_{i}+a_{i-1}}{2}\right)^{2}\cdot\vec{\textbf{c}}_{i}..\label{eq:Lift}
\end{equation}
The combined action of such longitudinal and lateral components (Fig.
\ref{Fig:lift-drag-forces}) provide the dissipative force term acting
on the $i-th$ bead

\begin{equation}
\vec{\textbf{f}}_{diss,i}=\vec{\textbf{f}}_{air,i}+\vec{\textbf{f}}_{lift,i}\:.
\end{equation}
Whereas, the random force term for the $i-th$ bead has the form

\begin{equation}
\vec{\textbf{f}}_{rand,i}=\sqrt{2m_{i}^{2}D_{\upsilon}}\cdot\vec{\boldsymbol{\eta}_{i}}(t),
\end{equation}
where $D_{\upsilon}$ denotes a generic diffusion coefficient in velocity
space (which is assumed constant and equal for all the beads), and
$\vec{\boldsymbol{\eta}_{i}}$ is a 3D vector, whereof each component
$\eta$ is an independent stochastic process, namely a nowhere differentiable
function with $<\eta\left(t_{1}\right)\eta\left(t_{2}\right)>=\delta\left(\left|t_{2}-t_{1}\right|\right)$,
and $<\eta\left(t\right)>=0$. Note that, for the sake of simplicity,
we assume $\eta=d\varsigma(t)/dt$, where $\varsigma(t)$ is a Wiener
process, namely a stochastic processes with stationary independent
increments (often called standard Brownian motion).\cite{durrett2010probability}

The sum of these forces governs the 
jet dynamics according to the Newton's equation providing the following non-linear Langevin-like
stochastic differential equation:

\begin{equation}
m_{i}\frac{d\vec{\boldsymbol{\upsilon}}_{i}}{dt}=\vec{\textbf{f}}_{el,i}+\vec{\textbf{f}}_{c,i}+\vec{\textbf{f}}_{\upsilon e,i}+\vec{\textbf{f}}_{st,i}+\vec{\textbf{f}}_{g,i}+\vec{\textbf{f}}_{diss,i}+\vec{\textbf{f}}_{rand}\:\text{,}\label{eq:force-EOM}
\end{equation}
where $\vec{\boldsymbol{\upsilon}}_{i}$ is the velocity of the $i-th$
bead. The velocity $\vec{\boldsymbol{\upsilon}}_{i}$ satisfies the
kinematic relation:

\begin{equation}
\frac{d\vec{\textbf{r}}_{i}}{dt}=\vec{\boldsymbol{\upsilon}}_{i}\label{eq:pos-EOM}
\end{equation}
where $\vec{\textbf{r}}_{i}\left(x_{i},y_{i},z_{i}\right)$ is the
position vector of the $i-th$ bead. The three Eqs \ref{eq:stress-ode},
\ref{eq:force-EOM} and \ref{eq:pos-EOM} form the set of EOM governing
the time evolution of the system. It is worth noting that Eq. \ref{eq:force-EOM}
recovers a deterministic EOM in the limit  $\rho_{a}$ and $D_{\upsilon} \rightarrow 0$  .

Furthermore, we define also the EOM of the nozzle bead located in
order to model fast mechanical perturbations at the spinneret.\cite{reneker2000bending,coluzza2014ultrathin}
Given the initial position of the nozzle $y_{n}^{0}=A\cdot\cos\left(\varphi\right)$
and $z_{n}^{0}=A\cdot\sin\left(\varphi\right)$ where $A$ and $\varphi$
are the amplitude and the initial phase of the perturbation, respectively,
the EOM for the nozzle bead are:

\begin{subequations}

\begin{equation}
\frac{dy_{n}}{dt}=-\omega\cdot z_{n}
\end{equation}

\begin{equation}
\frac{dz_{n}}{dt}=\omega\cdot y_{n},
\end{equation}

\end{subequations}

where $\omega$ denotes the perturbation frequency.
The actual perturbation at the nozzle produces a characteristic annular deposition 
of the fiber on the collector, as initially observed by Reneker et al. \cite{reneker2000bending}. 
Altough the collected fibers observed in experimental findings show less regular fiber patterns,
we find it convenient to investigate counterflow effects avoiding extra 
perturbations not directly related to the gaseous counterflow. 
Thus, we focus our investigation on the specific perturbation effect due 
to a counterflow gas on the jet dynamics.

Following previous works,\cite{Lauricella2015dissipative,lauricella2015nonlinear}
the EOM are integrated as follows. First, the time is discretized
as a uniform sequence $t_{i}=t_{0}+j\Delta t$, $j=1,\ldots,n_{steps}$.
At each time step and for each $i-th$ jet bead, we firstly integrate
the stochastic Eq. \ref{eq:force-EOM} using the explicit 
integration scheme proposed by Platen, \cite{platen1987derivative,kloeden1992numerical}
with  order of accuracy evaluated in literature
equal to $1.5$. Then, the Eqs. \ref{eq:pos-EOM} and
\ref{eq:stress-ode} are integrated via second order Runge-Kutta
integrator, where the $\vec{\boldsymbol{\upsilon}}_{i}\left(t+\varDelta t\right)$
value was previously obtained via the Platen scheme.

\section{Results and Discussion}

\subsection{Simulations setup for PVP electrified jets}

Solutions of polyvinylpyrrolidone (PVP) are largely used in electrospinning
experiments. In this work, we use a few simulation parameters developed
by Lauricella et Al. \cite{lauricella2015jetspin} and based on the experimental
data provided by Montinaro et Al. \cite{montinaro2015dynamics}. The process makes use
of a solution of PVP (molecular weight = 1300 kDa) prepared by a mixture
of ethanol and water (17:3 v:v), at a concentration ranging between
11 and 21 mg/mL. The relevant parameters include mass, charge density,
viscosity, elastic modulus, and surface tension, which were already
included in the model as implemented in JETSPIN.\cite{lauricella2015jetspin}
The extra parameters related to the gas environment are modeled on the
air (density $\rho_{a}=1.21\,\text{kg/m}^{3}$, and kinematic viscosity
$\nu_{a}=0.151\,\text{cm}^{2}/\text{s}$). The parameter
$D_{\upsilon,i}$ for the $i$-th bead is set to
be $\gamma_{i}$ for all the simulations. All
the $\gamma_{i}$ have the same value, and, consequently, $D_{\upsilon,i}$
is constant for all the beads. In addition, a perturbation is applied
at the nozzle with frequency $\omega=10^{4}\text{s}^{-1}$,
as proposed by Reneker et al.,\cite{reneker2000bending} whereas its
amplitude $A$ is equal to $0.01$ mm. The voltage bias between the
nozzle and the collector is 9 kV, and the collector is placed at 16
cm from the nozzle. The initial fluid velocity $\upsilon_{s}$ was
estimated considering a solution pumped at constant flow rate of 2
mL/h in a needle of radius $250\,$micron. For convenience, all the
simulation parameters are summarized in Table \ref{tab:simulation-param}.
We probe three different conditions of air flow velocity, $\upsilon_{flow}$.
In the first, we study the electrospinning process in absence of gas
flow, $\upsilon_{flow}=0$, which will be use as a reference case
(\textit{case I}). In the second and third, we take $\upsilon_{flow}=-10$
m/s (\textit{case II}), and $\upsilon_{flow}=-20$ m/s (\textit{case
III}), whose magnitudes are similar to the jet velocity measured at
the collector (about 20 m/s) in absence of gas streams. It is worth
stressing that the gas flow is oriented along the $x$-axis, and the
negative sign of $\upsilon_{flow}$ indicates its opposite direction
(counterflow, from the collector towards the nozzle). For each of
the three conditions, we run ten independent trajectories in order
to perform a statistical analysis. All simulations were carried
out by the modified version of the software package
JETSPIN,\cite{lauricella2015jetspin} and the corresponding EOM were
integrated with a time step of $10^{-9}$ seconds over a simulation span of 0.5 seconds.

For the sake of convenience, we report below the definition of few observables,
which will be used in the following. We define the jet length as:

\begin{equation}
\lambda\left(t\right)=\sum_{i=1}^{n-1}\left|\vec{\textbf{r}}_{i+1}-\vec{\textbf{r}}_{i}\right|
\end{equation}

with $\vec{\textbf{r}}_{i}$ the position vector, and $n$ the number
of jet beads. This observable takes note of the total length of the
jet from the collector up to the nozzle. Further, we introduce a suitable
observable to assess the tortuosity of the path, which is defined
as:

\begin{equation}
\Lambda\left(t\right)=\frac{\lambda}{\left|\vec{\textbf{r}}_{1}\right|},
\end{equation}

where $\left|\vec{\textbf{r}}_{1}\right|$ is the position vector
modulus of the closest bead to the collector. Note that $\Lambda$
tends to 1 for a rectilinear jet, and it takes larger
values depending on the complexity of the bending part of the jet.
We also define the instantaneous angular aperture of the instability
cone as:

\begin{equation}
\Theta\left(t\right)=\arctan\left(\frac{\sqrt{y_{1}^{2}+z_{1}^{2}}}{x_{1}}\right),
\end{equation}

with $x_{1}$, $y_{1}$ and $z_{1}$ the coordinates of the bead closest
to the collector (see Fig. \ref{Fig:model}).

In all the simulations, we observed two different regimes of the observables
$\left(\lambda,\Lambda,\Theta,\text{ etc.}\right)$ describing the
process. In the first stage, the jet has not yet reached the collector,
and we observe an initial transient of the observables. After the
jet touches the collector, the observables start to fluctuate around
a constant mean value, providing a stationary regime. As a consequence,
we discern two stages of the jet dynamics, hereafter denoted as early and late dynamics, respectively.

\subsection{Early dynamics}

\begin{figure}[H]
\begin{centering}
\includegraphics[scale=0.6]{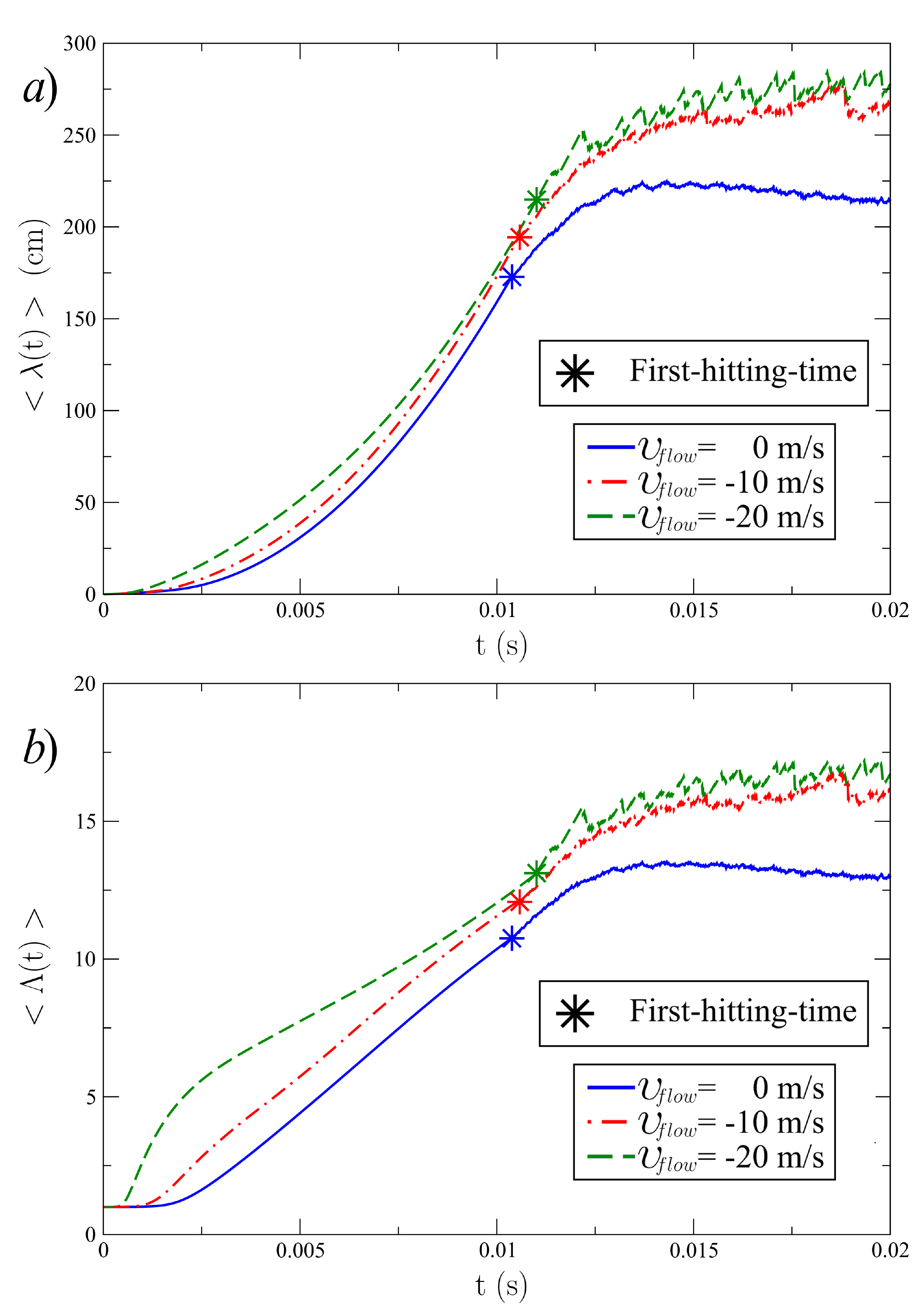} 
\par\end{centering}

\protect\caption{Time-dependent mean values of the observables jet length, $<\lambda\left(t\right)>$
(denoted $a$) and tortuosity degree, $<\Lambda\left(t\right)>$ (denoted $b$) for the different
cases of flow speed $\upsilon_{flow}$. Stars: times corresponding
to the mean value of the first-hitting-time, $<t_{first}>$, for each
case. }

\label{Fig:length-path} 
\end{figure}

\begin{figure}[H]
\begin{centering}
\includegraphics[scale=0.6]{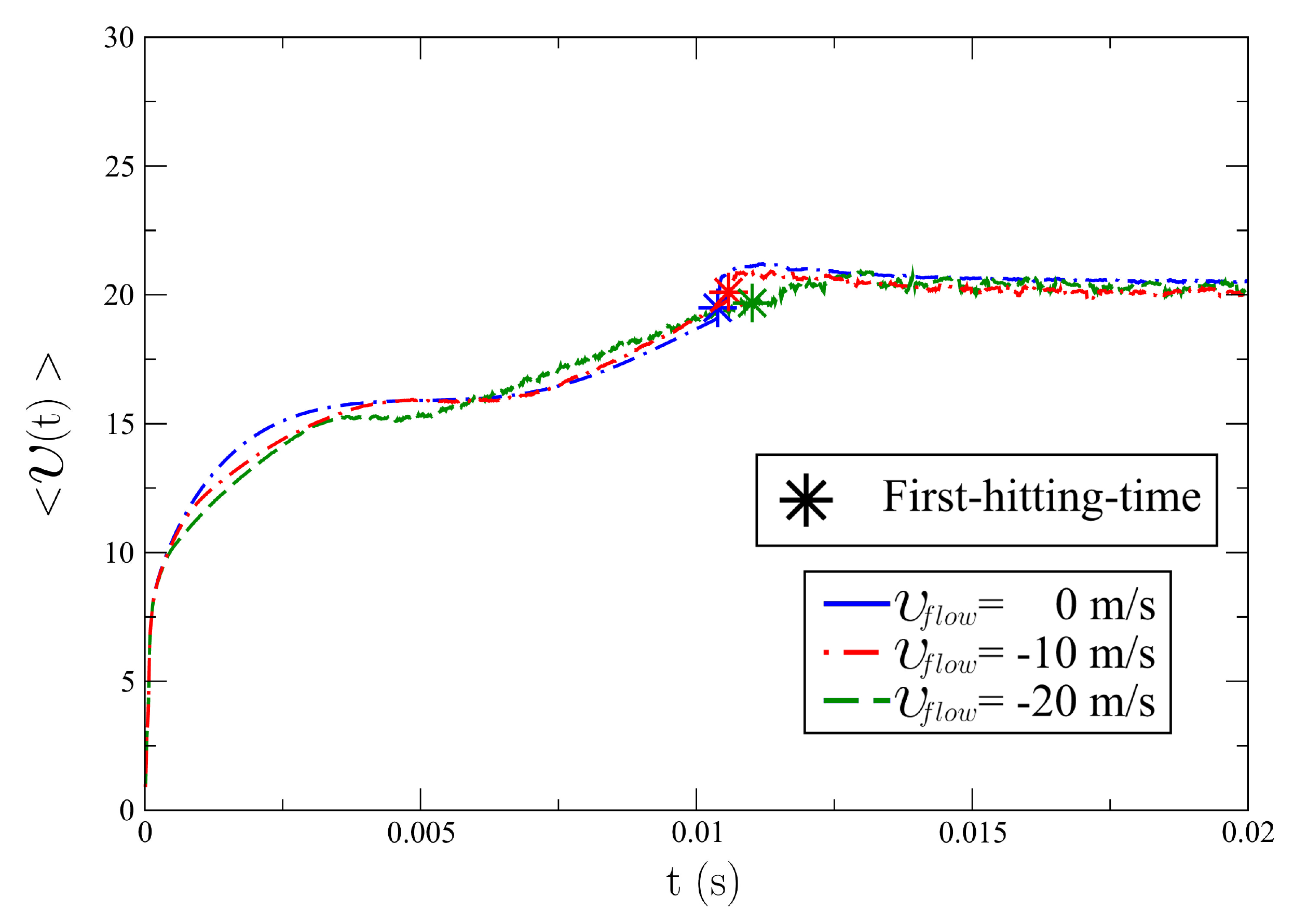} 
\par\end{centering}

\protect\caption{Time-dependent mean value of the jet velocity $<\upsilon\left(t\right)>$
(meter per second) as function of time (second) for all the three cases. Stars: times corresponding
to the mean value of the first-hitting-time, $<t_{first}>$, for each
case. }

\label{Fig:velocity-time} 
\end{figure}

For each case, we compute the average values of observables describing
the jet dynamics (see Fig. \ref{Fig:length-path}). The averages are
assessed at every step of the time integration, hence we obtain time dependent
mean values of observables along the jet evolution. In Table \ref{tab:mean-first-hitting}
we report the average first-hitting-time, $<t_{first}>$ , defined
as the time that the jet initially takes to touch the collector. In
particular, we note that the presence of a gas counterflow does not
affect significantly the first-hitting-time, and the velocity of the
jet bead at the collector is almost the same for all the three investigated
cases (within the margin of error). 
For the sake of completeness, we plot in Fig. \ref{Fig:velocity-time} the time dependent
mean velocity of the first bead as a function of time.
 On the other hand, a significant
increase of the jet length $<\lambda\left(t_{first}\right)>$ is found
upon increasing the gas counterflow speed $\upsilon_{flow}$. This
effect might be relevant for improving the quality of the resulting
fibers, since longer jet lengths usually correspond to smaller cross
sections of the deposited polymer filaments. Such increment of $<\lambda\left(t_{first}\right)>$
is due to the greater complexity of the jet path, where bending instabilities
play a significant role in determining the distance traveled from
the nozzle to the collector. This is well represented by the $\Lambda$
parameter, which increases by 20 \% in the \textit{case III}, when the
gas flow is set to $\upsilon_{speed}=-20$ m/s.

The dynamics of bending instabilities also deserves few comments:
we show in Fig. \ref{Fig:length-path} the time dependent mean value
of the jet length, $<\lambda\left(t\right)>$, and tortuosity degree,
$<\Lambda\left(t\right)>$, for each case under investigation. Here,
we find that bending instabilities start earlier for the \textit{case
III}, triggering a larger jet path in the subsequent dynamics. This
is well represented by the initial hump of $<\Lambda\left(t\right)>$,
which is already equal to 5.0 after 0.002 seconds. The
larger tortuosity degree is likely due to the lift force, which increases
the local curvature of the jet, as shown in Eq. \ref{eq:Lift}. Hence,
the synergic action of lift and Coulomb repulsive forces boost bending
instabilities at an earlier stage, and the \textit{case III} shows
a different dynamics, which is clearly evident in the initial 0.005
seconds. This effect substantially differs from what is reported in
literature for electrospinning models without external gas flows,
where only Coulomb repulsive forces contribute to the jet misalignement.\cite{reneker2000bending}

Furthermore, we note that $<\lambda\left(t\right)>$ increases for
all the cases both before and after the jet has touched the collector
for the first time, indicating that bending instabilities reach a
stationary regime of fluctuation at least after the time $t_{lim}\approx2\cdot<t_{first}>$.
We will consider this criteria in the following Subsection, in order
to discard the initial transient of dynamics for a correct statistical
analysis of the stationary regime.

\subsection{Late dynamics}

\begin{figure}[H]
\begin{centering}
\includegraphics[scale=0.7]{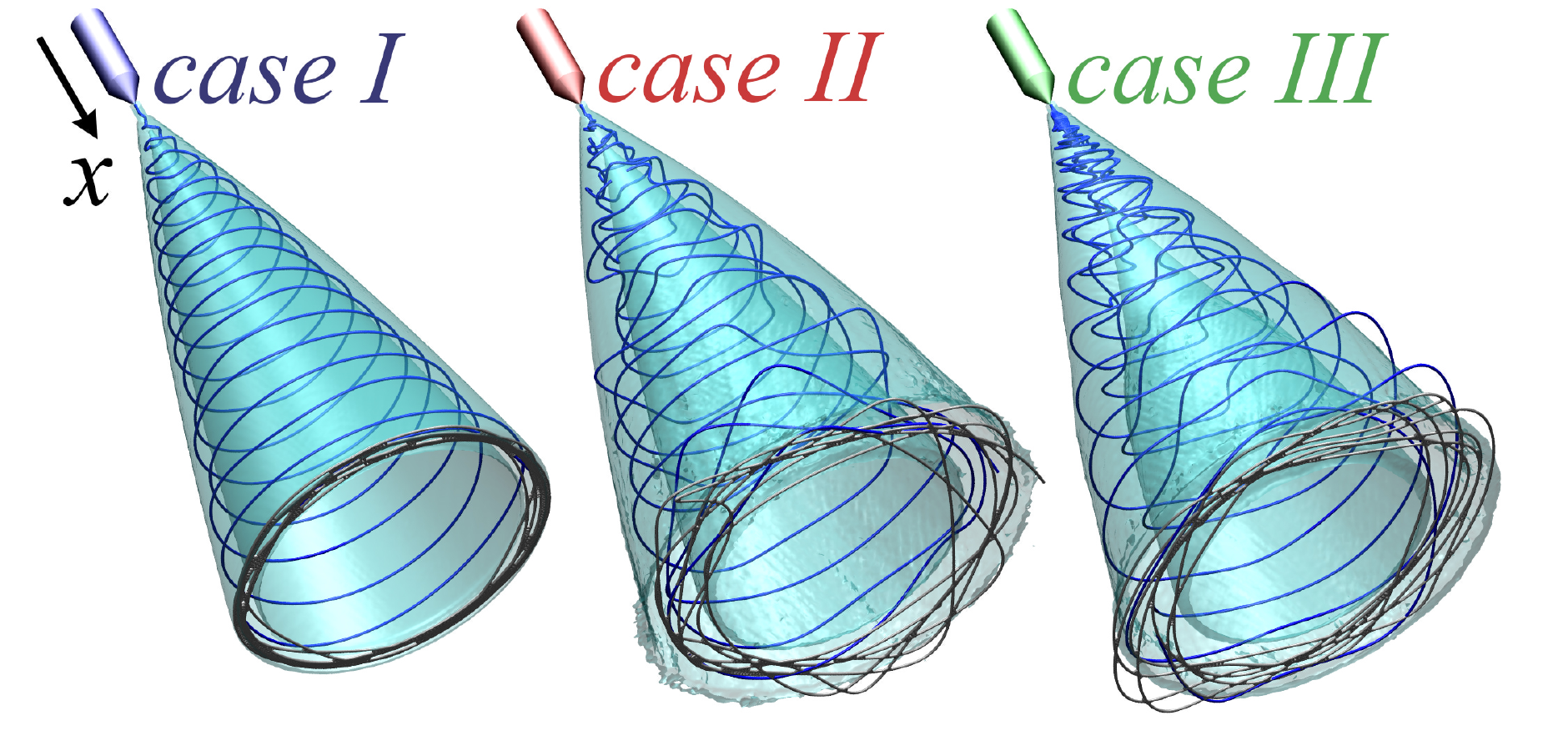} 
\par\end{centering}

\protect\caption{Simulation snapshots of the three different cases. From left to right
the snapshots correspond to the \textit{case I}, $\upsilon_{flow}=0$
m/s, \textit{case II}, $\upsilon_{flow}=-10$ m/s, and \textit{case
III}, $\upsilon_{flow}=-20$ m/s, respectively. The jet between the
nozzle and the collector is drawn in blue, and the fibers deposited
on the collector are colored in gray. The isosurfaces colored in cyan
represent the normalized numerical density field $\tilde{\rho}$ of
constant value equal to 0.001.}

\label{Fig:density} 
\end{figure}

\begin{figure}[H]
\begin{centering}
\includegraphics[scale=0.3]{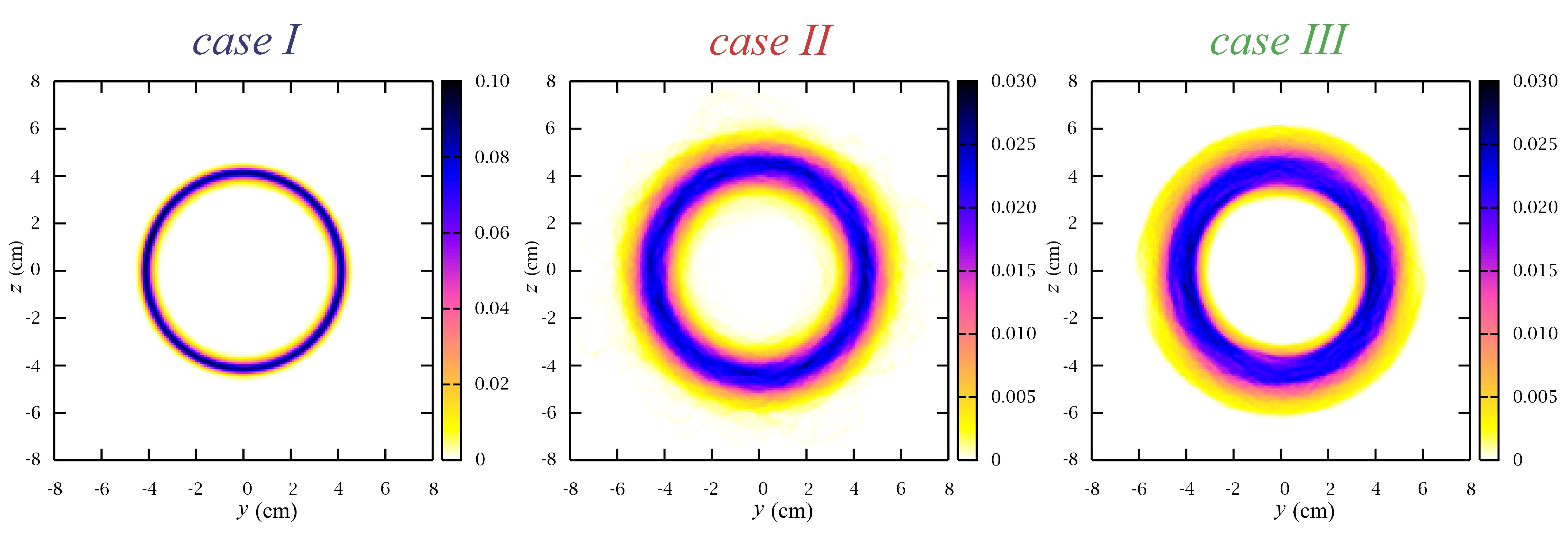} 
\par\end{centering}

\protect\caption{Normalized 2D maps computed over the coordinates $y$ and $z$ of
the collector for the three cases under investigation. The color palettes
define the probability that a jet bead hits the collector in coordinates
$y$ and $z$. }

\label{Fig:fist-collector} 
\end{figure}

We perform a statistical analysis of the positions of the jet beads
over all the ten independent simulations for each of the three cases
under investigation. In particular, we define an orthogonal box of
dimensions $16\text{cm}\times8\text{cm}\times8\text{cm}$ along the
$x$, $y$ and $z$-axes, respectively. The orthogonal box is discretized
in sub cubic cells of side equal to 1 mm, and the normalized numerical
density field, denoted $\tilde{\rho}_{i,j,k}$, is computed over all
the box for each case. By construction, $\tilde{\rho}_{i,j,k}$
provides the probability to find a jet bead in the cubic cell identified
by the indices $i$, $j$, $k$. As above, we discard the
initial part of each simulation, which corresponds to the early dynamics,
so that only the late dynamics describing the stationary regime is
considered. Hence, the dynamics of each trajectory is evolved in time
for 0.5 seconds. Fig. \ref{Fig:density} displays the isosurface of
$\rho_{n}$ representing points of constant value 0.001. The jet paths
statistically lie on an empty cone, whose aperture slightly increases
upon increasing the flow speed $\upsilon_{flow}$. In addition, the
chaotic behavior of jets is found to be enhanced by high-speed gas
flows. This is shown both by the larger statistical dispersion of
the cone (thickness of cone wall) and by the different shape of the
electrospun coatings deposited on the collector, which follow a fuzzier
path (gray fiber drawn in Fig. \ref{Fig:density}). The different
depositions of fibers for the three cases are highlighted by the normalized
2D maps in Fig. \ref{Fig:fist-collector}, where we show the probability
of a jet bead hitting the collector at the coordinates $y$ and $z$
(note that the plate is perpendicular to $x$ by construction). Here,
all the distributions are found to draw almost regular circles, which
subtend their relative instability cones of aperture angle $\Theta$.
The probability distribution of hitting a specific point on the collector
is remarkably peaked in the \textit{case I} without gas flow, whereas
the fiber deposition becomes less regular in the other cases. In particular,
the distributions lie within two concentric circles, whose inner
radius decreases, while the outer increases, as the air gas flow is
enforced. The trend is a consequence
of the more complex paths with highest tortuosity degree $\Lambda$ (see
\textit{case III} in Table \ref{Tab:mean-stationary}) drawn by the
jets under the effects of strong perturbation forces in presence of
a high speed gas counterflow. The snapshot related to the case
III in Fig. \ref{Fig:density} represents well the chaotic route followed
by the viscoelastic jet under the gas flow effects, which provides
a longer jet path length $\lambda$, whose mean value $<\lambda>$
increases by increasing the flow speed $\upsilon_{flow}$, as reported
in Table \ref{Tab:mean-stationary}. On the other hand, the mean values
of the aperture angle $\Theta$ are not significantly altered by the
gas flow (see Table \ref{Tab:mean-stationary}), showing that the
instability mainly alters the statistical dispersion of the cone,
but not in its mean value.

\begin{figure}[H]
\begin{centering}
\includegraphics[scale=0.6]{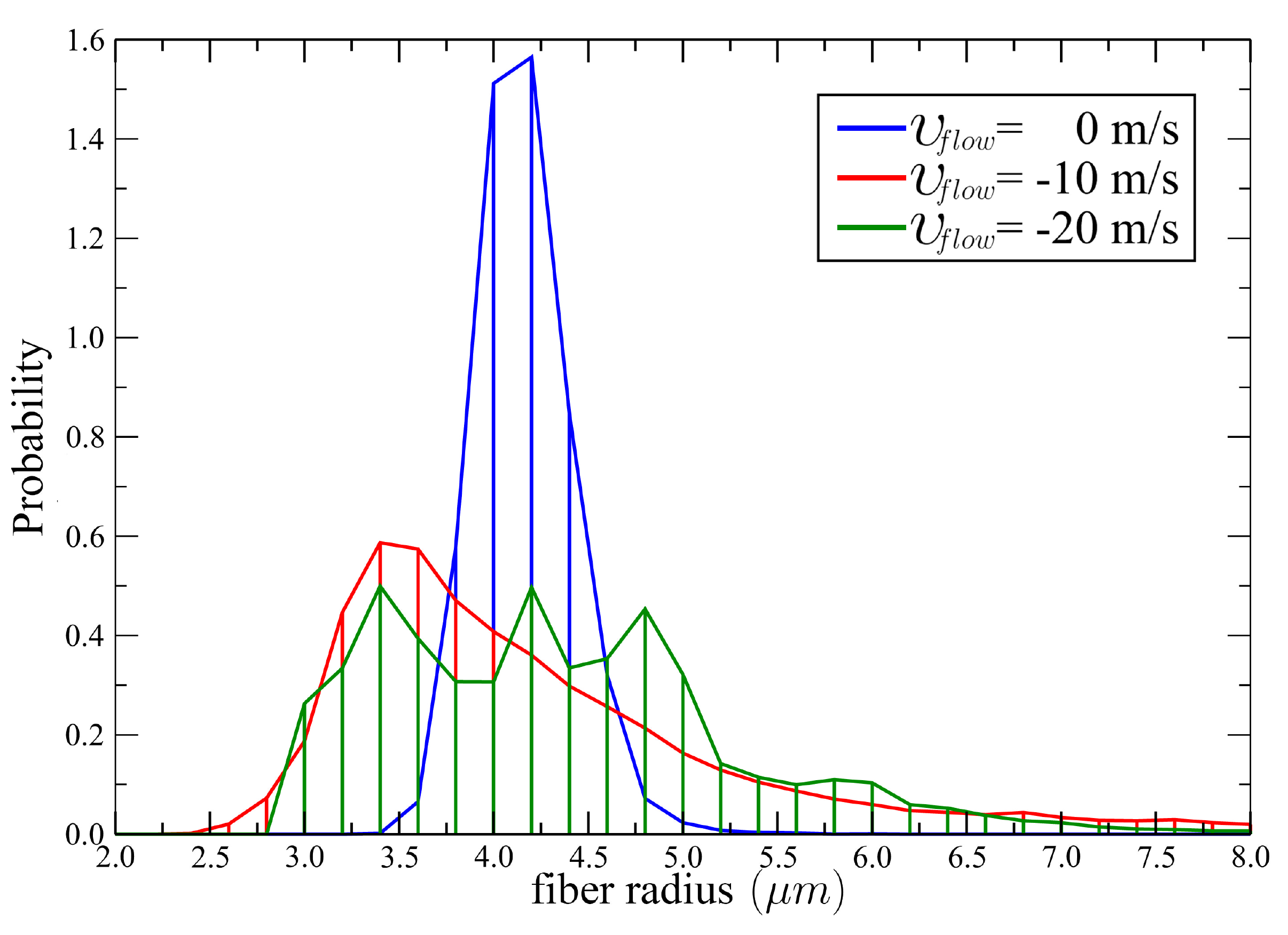} 
\par\end{centering}

\protect\caption{Normalized probability of depositing a fiber with a given radius.}

\label{Fig:hist-radius} 
\end{figure}

\begin{figure}[H]
\begin{centering}
\includegraphics[scale=0.4]{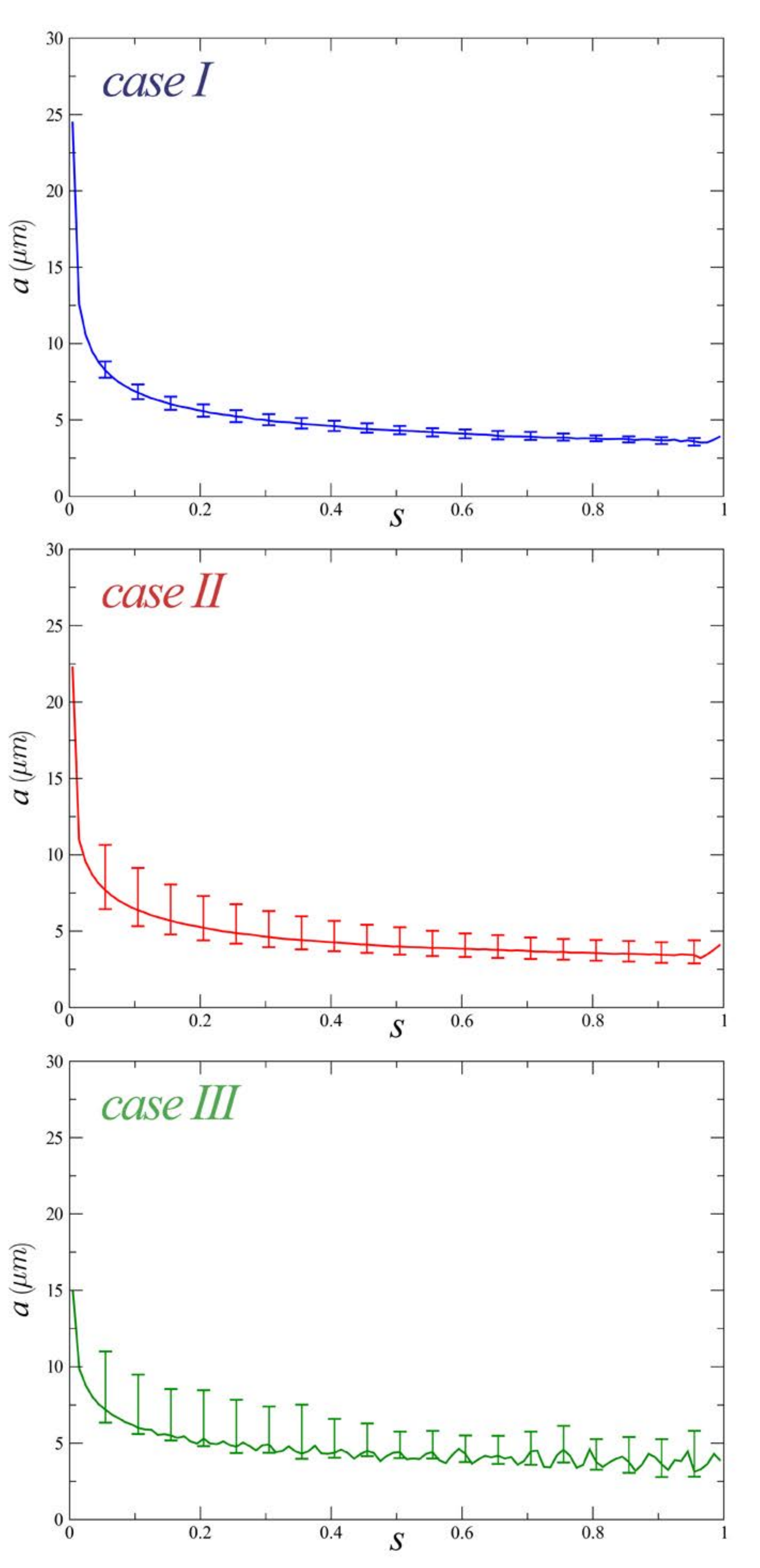} 
\par\end{centering}

\protect\caption{Meadian values of the jet radius distributions, $a$ (micrometer), computed along the curvilinear coordinate $s$ for all the three cases.
The error bars provide the amplitudes of the distributions evaluated as interquartile range.}

\label{Fig:dist-radius-curvlin} 
\end{figure}

The high-speed gas flow significantly affects the size distribution
of the deposited fibers. In Fig. \ref{Fig:hist-radius} we report
the probability of collecting fibers with a given value of cross-sectional
radius. Here, we observe a nontrivial trend of the fiber radius as
function of the air counterflow velocity. In particular, by applying
an air flow velocity $\upsilon_{flow}$ of -10 m/s (\textit{case II}),
we note a decrease in fiber radius by 10\%-15\%, and the fiber radius
probability distributions become broader. The latter effect is even
more evident for the \textit{case III} ($\upsilon_{flow}=$-20 m/s),
where the distribution computed over all the trajectories is spread
out from its mean with values of fiber radius oscillating between
$3$ and $8$ $\mu\text{m}$. Further, we observe a non-symmetric distribution of the fiber radius for both the 
\textit{cases II} and \textit{III}, which may appear somehow counterintuitive. 
Nonetheless, we wish to point out that skewed probability distributions are quite common in the statistical 
behavior of complex non-linear systems, such as the one considered here. 
Fluid turbulence is a typical example in point \cite{benzi1994scaling,ottaviani1990numerical}.
Although finding the coarse-grained dynamic equations of motion with respect to the jet cross section is beyond the aim of the present work, 
we investigate the phenomenon by computing the average distribution of the jet radius along the curvilinear coordinate $s$, where $ s \in[0,1]$ is 
introduced to parametrize the jet path, $ s=0$ identifies the nozzle, and $ s=1$
the filament at the collector. In Fig. \ref{Fig:dist-radius-curvlin} we report for all the three cases the median 
of the radius conditional distributions computed along the curvilinear coordinate $s$
(the condition is the given value of $s$).
We also report the amplitudes of the conditional distributions evaluated as interquartile range. 
Here, we observe that all the radius fluctuations are generated close to the nozzle.
In particular, at $s=0.05$ we already note non-symmetric fluctuations of the jet radius for the \textit{cases II} and \textit{III}.
Further, we observe larger average values of the curvature $k$ when the counterflow is activated. 
For instance, the averaged curvature measured
at $s=0.05$ is 1.1, 1.6 and 1.9 for the \textit{cases I}, \textit{II} and \textit{III}, respectively.
This is likely due to lift perturbation forces acting in junction with the Coulomb repulsive forces, which produce 
sharp bends along the jet path already close to the nozzle, providing large fluctuations in the jet cross section.
Thus, the quality
of the produced fibers is less controllable in presence of large counterflows
(as already evidenced in Fig. \ref{Fig:fist-collector} for the \textit{case
III}), and the beneficial effects of the gas stream in decreasing the fiber radius
are largely counteracted. Therefore, with the aim of producing thinner
fibers and at achieving narrower size distributions of the deposited
polymer filaments, the counterflow velocity $\upsilon_{flow}$ should
be carefully tuned, in order to provide an optimal balance between
dissipative and perturbation forces as related to the gas stream.

\section{Summary and Conclusions}

Summarizing, we have investigated the dynamics of electrified polymer
jets under different conditions of air drag force. In particular, we
have probed the effects of a gas flow oriented towards the nozzle
on the viscoelastic jet (counterflow) during the electrospinning process,
analyzing both the early and the late dynamics. Several observables
have been employed to analyze the air drag effects on the jet bending
instabilities, showing that the instability cone is altered in its
shape and aperture by the presence of a gas stream. Further, the results
in terms of fiber deposition were also investigated by a statistical
analysis of the late dynamics. We have observed that a controlled
gas counterflow might lead to a decrease of the mean value of the
fiber cross sectional radius. In particular, our data show a nontrivial
trend of the fiber radius as function of the air flow velocity applied
in electrospinning experiment. In fact, the gas flow generates both
dissipative and perturbation forces, which provide opposite effects
on the resulting fiber cross section. Thinner fibers are obtained
by using a gas flow speed of -10 m/s. The complex interplay of effects
due to air drag forces deserves a deeper investigation, which will 
be the subject of future work. However, further investigations 
will be needed and new terms have to be introduced to describe properly 
the disordered fiber structure experimentally observed on the collector. 
In particular, the effect of more complicated modeled perturbations of the nozzle in presence
of air counterflow could provide a more realistic pattern of the filament on the collector.
Anyway, the released model represents an important novelty and it might be used for
designing a new generation of devices with novel experimental
components for gas-assisted electrospinning, in order to further investigate
experimentally this process and to ultimately produce polymeric filaments
with finely controlled average diameters and size distribution.

\section*{Acknowledgments}

The research leading to these results has received funding from the
European Research Council under the European Union's Seventh Framework
Programme (FP/2007-2013)/ERC Grant Agreement n. 306357 (\textquotedbl{}NANO-JETS\textquotedbl{}).
The authors are grateful to Dr. G. Pontrelli for several useful discussions.

\newpage{}

\section*{Tables}

\begin{table}[H]
\begin{centering}
\begin{tabular}{cccccc}
\hline 
$\rho$  & $\rho_{q}$  & $a_{0}$  & $\upsilon_{s}$  & $\alpha$  & $\mu$ \tabularnewline
($\text{kg}/\text{m}^{3}$)  & ($\text{C}/\text{L}$)  & ($\text{cm}$)  & ($\text{cm/s}$)  & (N/m)  & (Pa$\cdot$s) \tabularnewline
\hline 
\hline 
840  & $2.8\cdot10^{-7}$  & $5\cdot10^{-3}$  & 0.28  & $2.11 \cdot 10^{-2}$  & 2.0 \tabularnewline
\hline 
\end{tabular}
\par\end{centering}

\begin{centering}
~ 
\par\end{centering}

\begin{centering}
\begin{tabular}{cccccc}
\hline 
$G$  & $V_{0}$  & $\omega$  & $A$  & $\rho_{a}$  & $\nu_{a}$\tabularnewline
(Pa)  & (kV)  & ($\text{s}^{-1}$)  & ($\text{cm}$)  & (kg/$\text{m}^{3}$)  & ($\text{cm}^{2}$/s)\tabularnewline
\hline 
\hline 
$5\cdot10^{4}$  & 9.0  & $10^{4}$  & $10^{-3}$  & 1.21  & 0.151\tabularnewline
\hline 
\end{tabular}
\par\end{centering}

\protect\caption{Simulation parameters for the simulations of electrified jets by PVP
solutions. The headings used are as follows: $\rho$: density, $\rho_{q}$:
charge density, $a_{0}$: fiber radius at the nozzle, $\upsilon_{s}$:
initial fluid velocity at the nozzle, $\alpha$: surface tension,
$\mu$: viscosity, $G$ : elastic modulus, $V_{0}$ : applied voltage
bias, $\omega$: frequency of perturbation, $A$ : amplitude of perturbation,
$\rho_{a}$ : air density, $\nu_{a}$ : air kinematic viscosity.}

\label{tab:simulation-param} 
\end{table}

\begin{table}[H]
\begin{centering}
\begin{tabular}{cccc}
\hline 
Observables  & \textit{case I}  & \textit{case II}  & \textit{case III}\tabularnewline
 & $\upsilon_{flow}=0$ m/s  & $\upsilon_{flow}=-10$ m/s  & $\upsilon_{flow}=-20$ m/s\tabularnewline
\hline 
\hline 
$<t_{first}>$ (s)  & $1.0385\cdot10^{-2}\pm8\cdot10^{-6}$  & $1.058\cdot10^{-2}\pm1\cdot10^{-5}$  & $1.101\cdot10^{-2}\pm2\cdot10^{-5}$\tabularnewline
$<\upsilon_{jet}\left(t_{first}\right)>$ ($\text{m}$/s)  & $19.6\pm0.2$  & $19.3\pm0.3$  & $19.5\pm0.4$\tabularnewline
$<\lambda\left(t_{first}\right)>$ ($\text{cm}$)  & $172.8\pm0.2$  & $194.4\pm0.7$  & $214.8\pm0.8$\tabularnewline
$<\Lambda\left(t_{first}\right)>$  & $10.8\pm0.1$  & $12.1\pm0.3$  & $13.1\pm0.4$\tabularnewline
\hline 
\end{tabular}
\par\end{centering}

\protect\caption{Mean values of the observables first-hitting-time $t_{first}$, and
mean values of the following observables at the first-hitting-time:
$\upsilon_{jet}\left(t_{first}\right)$ jet velocity measured at the
collector, jet path length $\lambda\left(t_{first}\right)$, and tortuosity
degree parameter $\Lambda\left(t_{first}\right)$. The averages were
computed over all the ten trajectories for each of the three cases
of gas flow speed $\upsilon_{flow}$. We report also the error as
standard deviation of distribution.}

\label{tab:mean-first-hitting} 
\end{table}

\begin{table}[H]
\begin{centering}
\begin{tabular}{cccc}
\hline 
Observables  & \textit{case I}  & \textit{case II}  & \textit{case III}\tabularnewline
 & $\upsilon_{flow}=0$ m/s  & $\upsilon_{flow}=-10$ m/s  & $\upsilon_{flow}=-20$ m/s\tabularnewline
\hline 
$<\Theta>$ (\textdegree )  & $28.1\pm1.2$  & $30.1\pm2.8$  & $29.6\pm2.9$\tabularnewline
$<\lambda>$ ($\text{cm}$)  & $213.8\pm2.2$  & $266\pm12$  & $279\pm13$\tabularnewline
$<\Lambda>$  & $13.4\pm0.1$  & $16.7\pm0.8$  & $17.5\pm0.9$\tabularnewline
\hline 
\end{tabular}
\par\end{centering}

\protect\caption{Mean values of the observables: aperture angle of instability cone
$\Theta$, jet path length $\lambda$, and tortuosity degree parameter
$\Lambda$. The averages were computed only in the stationary regime
over all the ten trajectories for each of the three cases of gas flow
speed $\upsilon_{flow}$. We report also the error as standard deviation
of distribution.}

\label{Tab:mean-stationary} 
\end{table}

\newpage{}

\section*{Figures}

\end{document}